# Photonic Topological Anderson Insulators

Simon Stützer[1], Yonatan Plotnik[2], Yaakov Lumer[3], Paraj Titum[4], Netanel Lindner[2], Mordechai Segev[2], Mikael C. Rechtsman[5], and Alexander Szameit[1]

[1]Institute for Physics, Rostock University, Albert-Einstein-Straße 23, 18055 Rostock, Germany

[2]Physics Department and Solid State Institute, Technion – Israel Institute of Technology, Haifa 32000, Israel

[3]Department of Electrical and Systems Engineering, University of Pennsylvania, Philadelphia, PA 19104, USA

[4]Joint Quantum Institute, NIST/University of Maryland, College Park, Maryland 20742 USA

[5]Department of Physics, Pennsylvania State University, University Park, PA 16802, USA

**The hallmark property of two-dimensional topological materials is the incredible robustness of the quantized Hall conductivity to disorder[6]. That robustness arises from the fact that in the topological band gap, transport can occur only along the edges modes, which are immune to scattering. However, for sufficiently strong disorder, the band gap closes and the system becomes topologically trivial as all states become localized, such that all transport vanishes- in accordance with Anderson localization[7,12]. It therefore came as a surprise when it was suggested that, for a two-dimensional quantum spin-Hall topological system, the opposite could occur. In so-called *topological Anderson insulators*[15], the emergence of protected edge states and quantized transport is caused by the introduction of disorder. However, to date, the observation of the topological Anderson insulator phase has been elusive. In this article, we report the first experimental demonstration of a topological Anderson insulator. We do that in a photonic implementation: an array of helical, evanescently-coupled waveguides in a detuned honeycomb geometry. Under proper conditions, adding on-site disorder, in the form of random variations in the refractive index contrast defining the waveguides, drives the system from a trivial phase into a topological state.**

The recent demonstration of photonic topological insulators[18–20] has shown that topological phenomena are not limited to the motion of electrons in solid-state materials. Rather, topological protection is a general wave phenomenon that applies equally well to many wave systems, including electromagnetic waves[18–29], acoustic waves[30,31], mechanical waves[32,33], and cold atoms[34,35]. Among these, photonic topological systems have been found to be useful in demonstrating effects that would be otherwise unavailable in the context of condensed matter physics, for example very strong strain[36,37], non-Hermitian behavior[38,39], and the concept of topological bound states in the continuum[40,41] (among others). Furthermore, topological photonic systems provide a complementary set of potential technological applications, including new mechanisms for integrated optical isolation and general robustness to photonic device fabrication imperfections.

Here, we demonstrate the first topological Anderson insulator. Our experiments are carried out in a photonic platform, as proposed theoretically[42], based on a time-reversal-broken Floquet topological insulator system. However, the key result is universal, and it specifically applies also to the BHZ model for the quantum spin Hall effect[15]. Namely, for certain parameters, adding disorder causes our system to pass from a topologically trivial regime, where there is no transport and all states are Anderson-localized, into a topological regime, where we observe unidirectional edge transport. The experimental platform we use is an array of evanescently-coupled helical waveguides, where the diffraction of light through the system is described by the paraxial wave equation, that is mathematically equivalent to the Schrodinger equation. A closely related system was used for the observation of Floquet photonic topological insulators[16].

To explain the mechanism underlying our photonic topological Anderson insulator (PTAI), we start with the honeycomb lattice of helical waveguides[18], a photonic Floquet topological insulator (see methods section for details of the mathematical description). Detuning the two triangular sublattices breaks the parity symmetry of the structure. The detuning opens a trivial bandgap, and can be quantified by the quantity $m_\delta$, interpreted as the mass associated with the quasiparticle governed by the effective Dirac equation of the honeycomb lattice in the absence of the periodic driving (helicity of the waveguides, in our photonic system). In the tuned case ($m_\delta = 0$) the Dirac cone dispersion is equivalent to that of massless relativistic particles. Such detuning can be realized experimentally in waveguide arrays by simply allowing the two honeycomb sublattices to have different refractive index contrasts.

The mechanism according to which we realize the PTAI is schematically depicted in Fig. 1. The band structure of the honeycomb lattice with waveguides which are straight and identical, such that the induced gauge field $A(z) = 0$ and $m_\delta = 0$[46], is shown in Fig. 1(a). This is a ribbon band structure (associated with the zig-zag edge, which is the termination along the left side of the lattices drawn in Fig. 1), which includes both edge and bulk bands. This is exactly the band structure of graphene[47], a two-dimensional material made of a honeycomb lattice of carbon atoms. As known in graphene physics, the ribbon band structure comprises two Dirac cones (red ellipses in Fig. 1(a)) that are connected by a flat band of edge states[48]. The valleys arise from Dirac cones, whose dispersion is equivalent to that of massless relativistic particles. When the waveguides follow a helical trajectory such that $A(z)$ is non-zero (Fig. 1(b)), the z-reversal symmetry is broken, and a band gap opens. In this case each valley acquires an opposite mass, namely $m_\tau$ and $-m_\tau$, respectively, for the left and right valleys. These opposite masses imply that the edge states cross the band gap, with each edge state localized to opposite sides of the structure. Therefore, the band gap is topological; the edge states form a single scatter-free chiral edge state that is localized to the perimeter of the structure. This is the essence of the Floquet topological insulator[18].

Consider now what happens when the sublattices of the Floquet topological insulator are detuned. Introducing non-zero detuning breaks the parity symmetry of the lattice, which adds a positive mass term $m_\delta$ to each valley. If the detuning is large enough ($m_\delta > m_\tau$), the masses $m_\tau + m_\delta$ and $-m_\tau + m_\delta$ in both valleys become positive and the band gap turns into a trivial one (Fig. 1(c)). In other words, the outcome of a sufficiently detuned honeycomb lattice of helical waveguides is a topologically trivial system with broken parity and time-reversal symmetry. This is where the disorder comes into play: as recently proposed[42] this system can be driven into a topologically non-trivial phase using disorder (Fig. 1(d)). More specifically, it was shown[42] (using the Born approximation) that introducing on-site disorder (i.e., randomizing the refractive index contrasts of the waveguides) causes an effective decrease of the detuning $m_\delta$. The decrease in $m_\delta$ grows stronger as the disorder strength is increased. Thus, upon introducing disorder and increasing its strength, the band gap closes when $m_\tau = m_\delta$ and reopens for $m_\tau > m_\delta$.

When the gap reopens, the system is topological (Fig. 1(e)). This is precisely the PTAI effect: adding disorder brings the system from being topologically trivial ($m_\delta > m_\tau$) to topologically non-trivial ($m_\tau > m_\delta$).

The schematic depiction of the waveguide configuration is shown in Fig. 2(a). In this setting, it is essential to control which states are excited. We do that by adding a one-dimensional auxiliary array - a "straw" - through which we selectively excite the modes of the system at a fixed energy (i.e., $k_z$). More specifically, in a 1D waveguide array, tilting the phase front of the wave illuminating the input facet determines the momentum of the excited Bloch mode[49]. Here, such selective excitation is especially important because we would like to excite states entirely within the gap, to determine whether the gap is trivial or topological. By controlling the tilt angle of input of the beam, we can control $k_x$, the Bloch wavevector in the horizontal direction. The band structure ($k_z$ vs. $k_x$) of the straw is shown in the inset in Fig. 2(a). We fix $k_z$ (the Floquet quasienergy) by choosing $k_x$, as $k_z$ is conserved even as the beam travels through the straw and enters the honeycomb array. Thus, simply by changing the incidence angle of the beam, we can tune through the band structure of the honeycomb lattice. This is the photonic analogue of tuning the "Fermi level" in a solid-state system. The bandwidth in $k_z$ of the straw depends on $k_x$ and is relatively wide (determined by the number of sites in the straw and fabrication precision), limiting our ability to perform precise spectroscopy. However, it remains a highly useful tool to control the quasienergy of the modes excited in the honeycomb lattice.

We first consider a system that breaks z-reversal symmetry ($m_\tau \neq 0$) with identical (non-detuned) waveguides ($m_\delta = 0$), and essentially repeat the previously realized Floquet topological insulator[18]. Specifically, we use a helix pitch of $Z = 1cm$ and radius of $R = 10\mu m$; at our probe wavelength ($\lambda = 633nm$), this corresponds to a dimensionless gauge field strength of $|A| = kR\Omega a = 1.26$ (with $k = 2\pi n_0/\lambda$; $\Omega = 2\pi/Z$, ambient refractive index $n_0 = 1.45$ and $a = 14\mu m$ nearest-neighbor spacing). Due to the helicity of the waveguides, the system has a topological band gap located at quasienergy $k_z = 0$, and we expect to find a chiral edge state in this band gap (the Floquet band structure is shown in the inset to Fig. 2(a)). This is precisely what we see: Fig. 2(b) shows the output facet of the array in the case where $k_z$ lies in the band gap (at top), and the case when it lies in the band (at bottom). When $k_z$ is in the gap, a chiral edge state is excited: the optical wavepacket is launched from the straw, it couples to the chiral edge state, where is propagates unidirectionally upwards in a clockwise direction around the honeycomb lattice, but not penetrating into the lattice itself. On the other hand, when $k_z$ lies in the band - the wavepacket couples to bulk states, hence it penetrates into the lattice and spreads throughout the array, while not staying confined to the edge. The evolution of the edge state is shown in Fig. 2(c): as the input beam is brought closer to the array, the edge state travels farther along the edge until it passes the top corner. The fact that it moves upwards only, and stays confined to the edge is a signature of the chirality of the edge state. We note that for the choice of parameters of the waveguide array used in the experiment, the spatial frequency of the helix $\Omega$ is smaller than the total bandwidth ($6c$) in the absence of the helix. This results in an additional topological gap, which opens around quasienergy $k_z = \pi/\text{cm}$, and hosts chiral edge states, as shown in Fig 2(a). However, by using the selective excitation through the straw, in this experiment we always excite states whose quasienergies are close to $k_z = 0$, and do not probe states with quasienergy near $k_z = \pi/\text{cm}$.

We now fabricate another photonic lattice, where we introduce detuning between the sublattices ($m_\delta \neq 0$) of the honeycomb structure by making their refractive index contrasts different. In practice, this is done by changing the speed of the laser-writing beam during the fabrication process (higher writing speed results in a lower refractive index of the waveguides). In a series of six waveguide arrays, we systematically increase this detuning, decreasing the gap size as $m_\delta$ increases, and examine how much the wave, launched through the straw, penetrates into the lattice (plotted in Fig. 3). To do that, we choose $k_x = \pi/2a$, where the Bloch wavenumber in the straw corresponds to the $k_z$ in the center of the band gap of the honeycomb lattice. For sufficiently strong detuning $m_\delta$ we observe the emergence of a sharp decrease in penetration depth into the array. This corresponds to the closing of the topological and reopening of the trivial band gap at $k_x = 0$, and this constraints $m_\delta$ to be larger than 1.5/cm). Indeed, in the trivial gap, no states whatsoever are present neither in the bulk nor at the edge, hence there is no coupling to bulk modes. This situation corresponds to launching a wave directly at an energy at the center of the bandgap of a photonic crystal, where no wave can penetrate and the light is completely reflected. The inability to penetrate into the lattice, and especially the inability to couple to edge states, establishes that we have introduced sufficiently strong detuning (sufficiently large $m_\delta$) to have opened the trivial band gap, which does not support edge modes.

Next, we introduce disorder, and prove the formation of the PTAI, namely, we add on-site disorder and observe whether the mid-gap excitation allows coupling to the topological edge state. The disorder enters the right hand side of Eq. (1) on the right-hand side with the additional term, $wr_i\psi_i$, where $r_i$ is a uniformly-distributed random variable between $-0.5$ and $0.5$, and $w$ is the disorder strength. We find that for sufficiently strong disorder (corresponding to a maximum variation in laser write-speed of *8mm/min*), the mid-gap excitation is able to couple into the lattice, including partially to the bulk, but overall light stays largely confined to the edge (see Fig. 4(a-c)), i.e., populating the edge state. As we move the input beam closer to the array, the light coupled to the edge state propagates further along the edge, much as in the non-disordered case of Fig. 2(c). We find the group velocity of the edge state to be 21μm/cm. The beam moves up along the left edge, implying the presence of a chiral edge state (that was not present when the system was not disordered). We emphasize that the light moves along the edge in a unidirectional fashion: only up, never down. For comparison, we show the same scenario, but with no disorder, in Fig. 4(d-f). Clearly – when no disorder is present, there is no observable edge state, and minimal bulk penetration, i.e., the beam launched into the topologically trivial photonic bandgap is reflected. The small amount of bulk penetration seen in Fig. 4(d-f) likely arises from the non-zero linewidth of the input beam in $k_z$. The observation of the appearance of the chiral edge state when the disorder is sufficiently strong is the smoking gun of the PTAI phase[42]. Figure 4(g) shows a phase diagram for the system as a function of detuning, $m_\delta$, and disorder strength, $w$. The phase diagram is obtained by calculating the Bott index of the Floquet band[42,51] which lies in the quasienergy interval $0 \leq k_z \leq \pi$/cm. The Bott index essentially counts[42] the number of chiral edge states traversing the gap at $k_z = \pi$/cm minus the number of chiral edge states traversing the gap at $k_z = 0$. The gap at $k_z = \pi$/cm hosts a single chiral edge states throughout the entire range of parameters shown; therefore, a Bott index of 0 corresponds to a topological gap at $k_z = 0$ and a Bott index of 1 corresponds to a trivial gap at $k_z = 0$. Figure 4(g) shows the Bott index averaged over an ensemble with fixed disorder strength $w$ [52,53]. The phase diagram clearly shows that, by adding sufficiently strong disorder, we necessarily take the system from the trivial into the topological phase. Naturally, for considerably stronger disorder, localization sets in, rendering the system Anderson localized[15].

The experiments presented here demonstrate the manifestation of a topological Anderson insulator, using a photonic platform. The TAI phase is a clear example of the complex relationship between topology and disorder in two dimensions: it goes beyond the notion that in topological systems, conduction is simply quantized until disorder is sufficiently strong to close the band gap[47]. It shows that disorder is a key variable in probing phases of electrons and light, in the sense that even in seemingly trivial systems, topological behavior can emerge when disorder is introduced at sufficient strengths. We expect that our experimental realization of this phase will stimulate a range of new theoretical and experimental studies exploring the role of disorder in tuning between topological phases. This prompts various important questions: can the topological Anderson insulator phase be realized in optical lattices of ultracold atoms, where topological quantities can be observed via bulk wavepacket dynamics, rather than by chiral edge states? Furthermore, what happens to the topological Anderson insulator phase in the presence of interactions, whether in optical, atomic, or condensed matter systems? Is there a topological Anderson insulator phase in the quantum many-body regime? The answers to these questions are now in experimental reach.

**Data availability statements**

All data generated or analysed during this study are included in this published article (and its supplementary information files).

**Author Contributions**

All authors contributed considerably to the work.

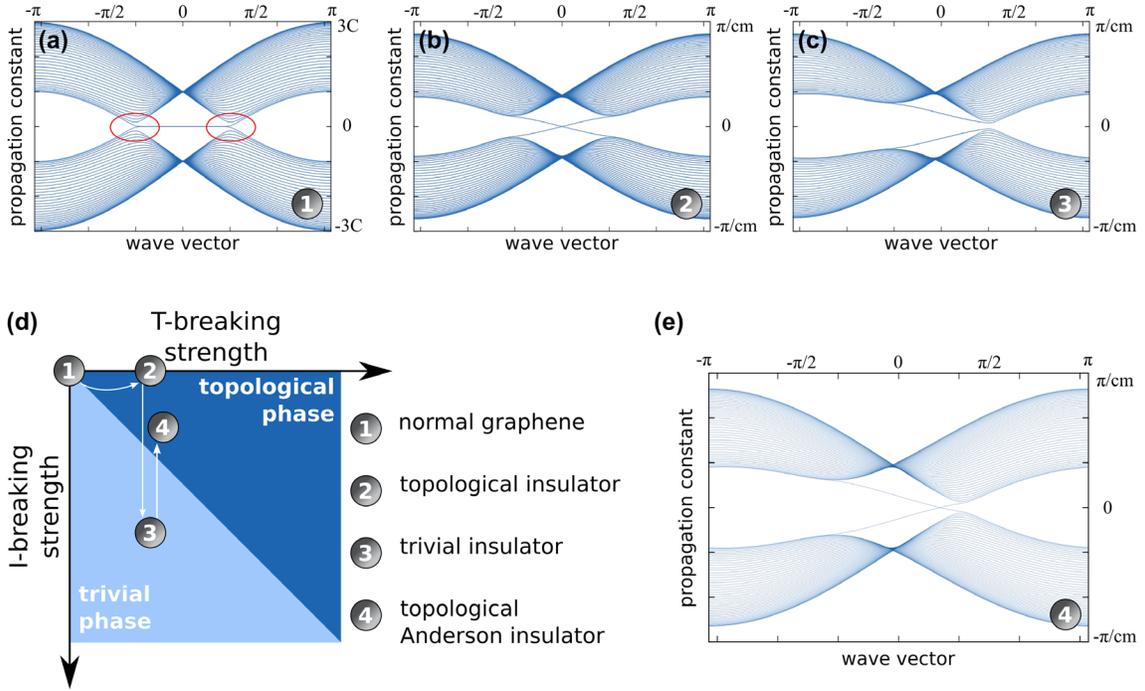

Figure 1: Mechanism of realizing the topological insulator phase in a photonic waveguide lattice. The parameters for the band structure calculations are chosen for illustration purposes. In a trivial honeycomb lattice with straight waveguides (a), the time-reversal symmetry is broken by helical waveguide trajectories, such that a topological band gap opens (b). Breaking the parity of the structure by sufficiently detuning the sublattices, a trivial band gap forms (c). Sufficiently strong disorder suppresses the effect of the parity breaking terms and moves the system into the topological Anderson insulator phase (d). Panel (e) shows the band structure corresponding to the renormalized parameters in the topological Anderson insulator phase, as obtained from the Born approximation (Please note that this figure is a pedagogical description: there is no well-defined band structure in the disordered case).

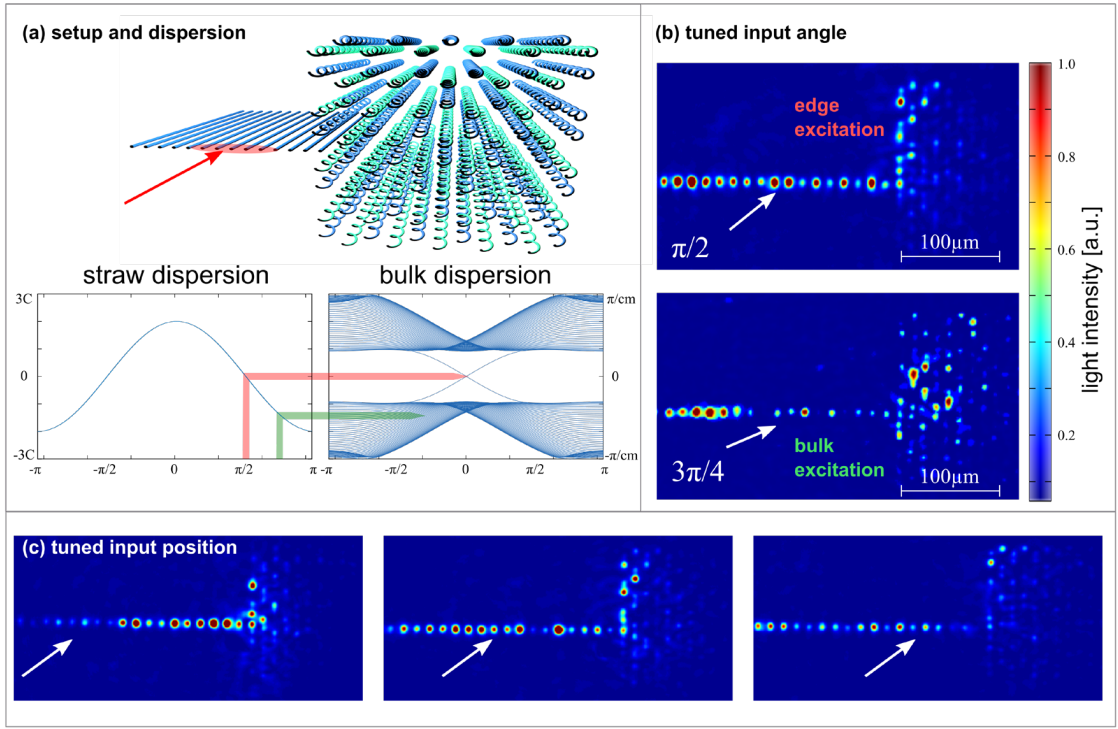

Figure 2: Setup and functionality of the lattice system. (a) Hybrid structure composed of a one-dimensional straw and a two-dimensional honeycomb lattice of helical waveguides with the corresponding dispersion relations. (b) Determining the correct input angle for the incident light such that a chiral edge states in the topological gap is excited. If the launch angle does not correspond to excitation in the gap, extensive coupling into the lattice bulk is observed. (c) Changing the input position facilitates observing the evolution of the edge state as they propagate along the edge and pass the upper corner, never penetrating into the bulk.

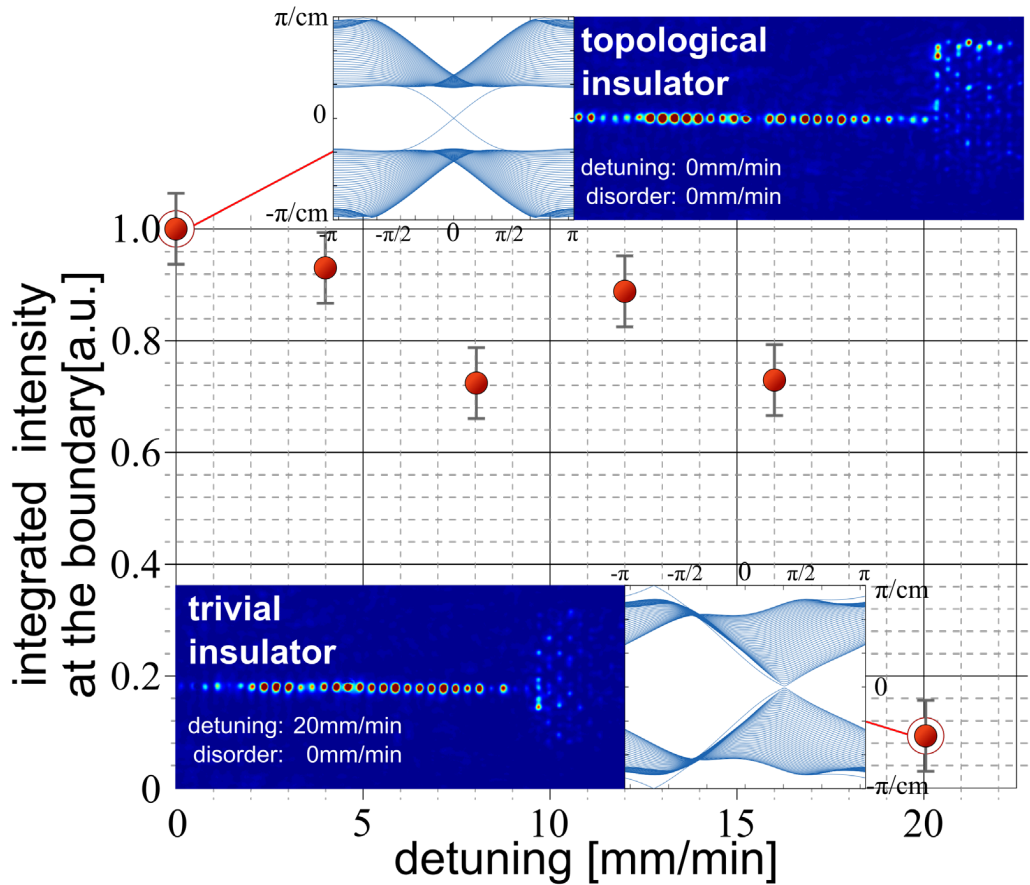

Figure 3: Engineering the topologically trivial phase. For sufficiently strong detuning between the sublattices of the honeycomb a trivial gap opens, such that the chiral edge state ceases to exist and all incident light is reflected back into the straw. The experiments show the light intensity pattern as it exits the helical waveguide array, and the measured integrated intensity at the boundary. For zero detuning, the exiting light resides only on the edge (hallmarking the topological insulator phase). When the detuning is small, the system is still in the topological phase and almost all the exiting light resides on the edge. When the detuning is further increased – the system becomes topologically trivial, and the incident light is entirely reflected.

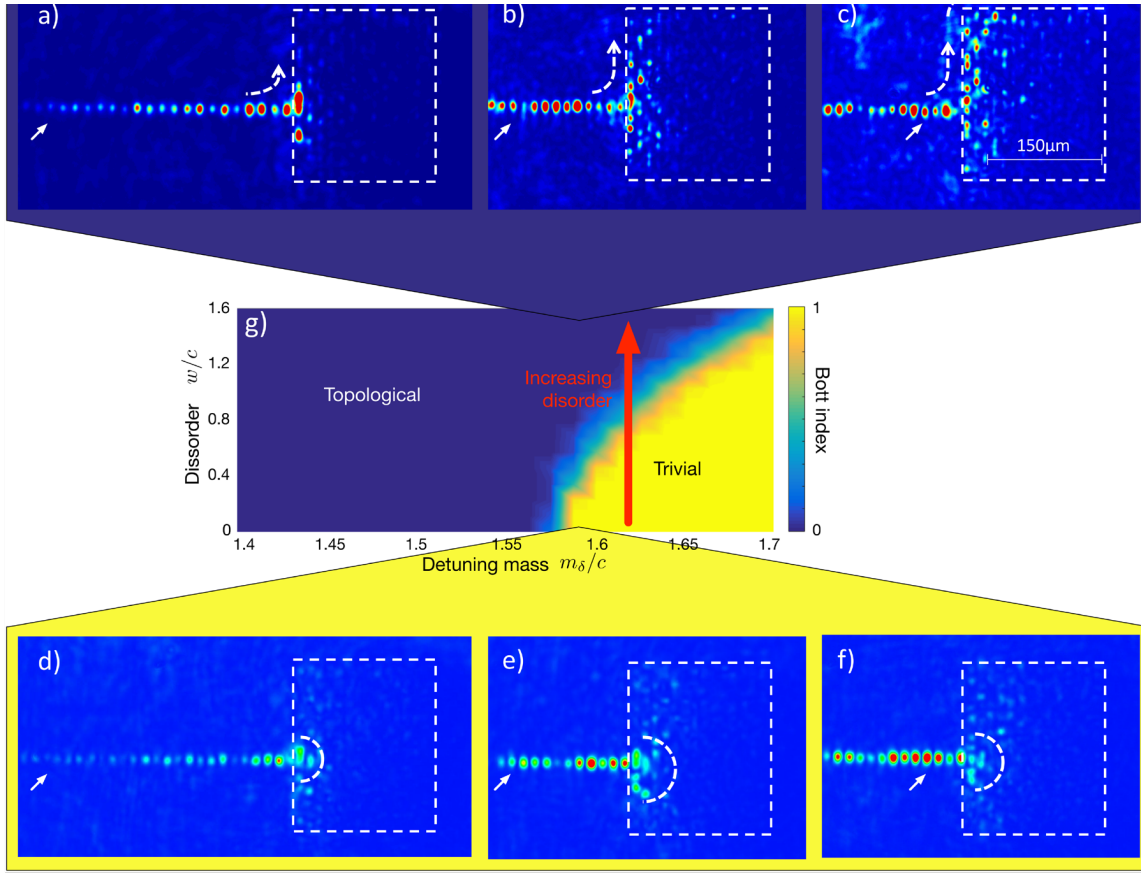

Figure 4: Formation of the PTAI. When adding sufficient disorder, the system is driven into the topological Anderson insulator phase and chiral edge states form. This is proven by moving the excitation in the straw close to the (disordered) honeycomb lattice, such that the edge state crawls up the edge and travels even around the corner (a-c). In comparison, in the fully ordered detuned helical honeycomb lattice essentially all light is reflected into the straw, with only a small part penetrating into the bulk without exciting any edge mode (d-f). A phase diagram showing the trivial and topological phases is shown in (g) as a function of the detuning mass $m_\delta$ and disorder strength, $w$. The phase diagram is established by a calculation of the Bott index[42,51-53] (a Bott index of 1 corresponds to a trivial gap at $\varepsilon = 0$ while a Bott index of 0 corresponds to topological gap $\varepsilon = 0$, see text). The red arrow indicates the trajectory through the phase diagram as the system starts from the trivial phase and enters the topological phase with increasing disorder strength.

**Methods**

Honeycomb photonic lattice with helical waveguides:

The equation describing the diffraction of a paraxial beam of light in this lattice can be written, under the tight-binding approximation, as:

$$i\partial_z \psi_i = c \sum_{<i,j>} e^{-i\boldsymbol{A}(z)\cdot \boldsymbol{r}_{i,j}} \psi_j + m_\delta \delta_i \psi_i \equiv \hat{H}(z)\psi, \qquad (1)$$

where $z$ is the distance of propagation along the waveguide axis; $\psi_i$ is the envelope function of the electric field in the $i^{th}$ waveguide; $c$ is the coupling strength between waveguides; $\boldsymbol{A}(z) = kR\Omega a(cos\Omega z, sin\Omega z, 0)$ is the gauge field induced by the helicity; $k$ is the wavenumber of the light in the medium (fused silica); $R$ is the helix radius; $\Omega$ is the spatial frequency associated with the helix; $a$ is the nearest-neighbor spacing; $\boldsymbol{r}_{n,m}$ is the displacement vector pointing from waveguide $i$ to $j$. The honeycomb lattice comprises two triangular sublattices. The parameter $\delta_i$ takes the value *1* in one sublattice and *-1* in the other, such that the on-site energies of the two are separated by the detuning $2m_\delta$. Equation 1 defines $\hat{H}(z)$ as the Hamiltonian at propagation distance *z*, and the summation therein is taken over nearest-neighbor waveguides. This is exactly the Schrödinger equation, where *z* takes the role of the temporal (time) coordinate. Since $\boldsymbol{A}(z)$ is *z*-dependent and periodic, the solutions to Eq. (1) can be obtained by employing Floquet's theorem. Thus, the band structure can be obtained by diagonalizing the unitary evolution operator for one period[M1-M3].

Details of the fabricated lattice:

The spacing between adjacent waveguides is *a=14μm*, which results in a coupling strength *c=1.8cm⁻¹*. The waveguides are written using the laser direct write technology[M4] with an average writing speed of 90mm/min. A broad input beam (wavelength *633nm*) is incident on the straw, with a beam waist of ~100μm.

Decay length of the edge state:

We extract the decay length of the edge state from the experimental data. To achieve this, we studied the intensity profile of the light propagating along the edge. To average over the disorder potential, we integrated the intensity per pixel along a direction parallel to the edge, and obtained the averaged intensity as a function of distance from the edge. The intensity profile, shown in Supplementary Fig. 1(a), decays with increasing distance to the edge, with peaks at the waveguide positions. We fit the envelope of the intensity profile to an exponential decay which results in a decay length of the edge state of 24 µm, see Supplementary Fig. 1(b). This length scale is thus much (more than 6 times) smaller than the system size, which is 158 µm. This result demonstrates that the experimental system in which we observed the topological phenomena observe is robust even in though the system has a finite size, as any experimental system must be system. We emphasize that we do not observe any light at the far edge of the sample (the edge opposite to the point where the straw meets the sample), which demonstrates that the edge decay length, as well as the bulk localization length at the energies we are probing, are strongly localized.

We compared the experimentally obtained value of the decay length of the edge states with theoretical values obtained from a numerical simulation of tight binding model. To extract the decay length from the numerical simulation, we numerically computed the Green's function $G_N(\boldsymbol{r_0}, \boldsymbol{r}, NT) = \langle \boldsymbol{r}|U(t = NT, 0)|\boldsymbol{r_0}\rangle$, evolved over $N$ periods. From the Fourier transform of the Green function in time, we computed $g_N(\boldsymbol{r_0}, \boldsymbol{r}, \varepsilon) = \langle |G_N(\boldsymbol{r_0}, \boldsymbol{r}, \varepsilon)|^2 \rangle$, where the brackets denotes disorder averaging [for more details on the definition of these quantities[42]. For initial positions $\boldsymbol{r_0}$ localized on the edge, and for epsilon in the mobility gap (near quasienergy $\varepsilon = 0$), the function $g_N(\boldsymbol{r_0}, \boldsymbol{r}, \varepsilon)$ shows propagation along the edge and an exponentially localized profile confined from the edge. We integrated $g_N(\boldsymbol{r_0}, \boldsymbol{r}, \varepsilon = 0)$, averaged over 100 realizations, along the direction parallel to the edge. We extract the decay length from the decay profile of the result (which is only a function of the distance from the edge). The results, shown in Supplementary Fig. 1(c) give a decay length of 6.5 lattice constants, which is in reasonable agreement with the experimental result.

From the numerically obtained disorder averaged Green's function, we also extract the localization length of the bulk. We take an initial position $r_0$ in the bulk of the system, and plot the corresponding function $g_N(\boldsymbol{r_0}, \boldsymbol{r}, \varepsilon = 0) = \langle |G_N(\boldsymbol{r_0}, \boldsymbol{r}, \varepsilon)|^2 \rangle$, see Supplementary Fig. 1(d). This yields a bulk localization length of 4 lattice constants. Our numerical analysis thus shows that the bulk localization length is indeed much smaller than the system size, which guarantees well defined edge states and topologically protected edge transport.

We note that the numerical simulation is based on a tight binding model, which is only an approximation to the actual dynamics of light in the optical system. Thus, we cannot expect perfect agreement between the numerical results from the tight binding model and the numerical data. Still, the length scales we extract from the experiment and the numerical simulation are in qualitative agreement, and the edge localization length in the optical system should be determined directly from the experimental data, as we have done above.

Spatial frequency resolution of the experimental system:

The plot in Supplementary Fig. 2 shows the integrated intensity at the boundary of the lattice as a function of the excitation angle in the straw for the disordered-detuned state in the topological Anderson phase, indicating the presence or absence of an edge state. For $k = \pi/2$ the formation of the edge state is clearly visible. The plot in Supplementary Fig. 3 shows the integrated intensity in the bulk in the trivial phase as a function of the excitation angle in the straw. Here we see that bulk penetration is at its minimum within the band gap ($k = \pi/2$), indicating a trivial band gap.